\documentclass[
onecolumn
,notitlepage
,aps
,pra
,11pt
,superscriptaddress]{revtex4-1}
\usepackage[utf8]{inputenc}
\usepackage{braket}
\usepackage{nicefrac}

\usepackage{amsthm,amsmath,amssymb,amsfonts}
\usepackage{mathtools}
\usepackage{latexsym}
\usepackage{graphicx}  
\usepackage{bm}        
\usepackage{verbatim}
\usepackage{soul}
\usepackage{mathpazo}
\usepackage{enumerate}
\usepackage{comment}
\usepackage[caption=false]{subfig}
\usepackage[usenames,dvipsnames]{xcolor}
\usepackage{tikz}
\usepackage[framemethod=tikz]{mdframed}
\usepackage[pagebackref=false]{hyperref}
\usepackage[nameinlink,capitalize]{cleveref}
\theoremstyle{plain}

\usepackage{tikz}
\usepackage{pgfplots}
\usetikzlibrary{plotmarks}

\newlength\figureheight
\newlength\figurewidth
\pgfplotsset{compat=1.14, 
             every axis/.append style={
                    tick label style={/pgf/number format/fixed}
                    }}

\providecommand{\id}{\hat{\mathbb{1}}}

\begin{document}
\title{Decoding Small Surface Codes with Feedforward Neural Networks}
\author{Savvas Varsamopoulos}
\author{Ben Criger}
\author{Koen Bertels}
\affiliation{Computer Engineering, Delft University of Technology, Mekelweg 4, 2628 CD Delft, The Netherlands}
\affiliation{QuTech, Delft University of Technology, P.O. Box 5046, 2600 GA Delft, The Netherlands}
\date{\today}
\begin{abstract}
Surface codes reach high error thresholds when decoded with known algorithms, but the decoding time will likely exceed the available time budget, especially for near-term implementations.
To decrease the decoding time, we reduce the decoding problem to a classification problem that a feedforward neural network can solve.
We investigate quantum error correction and fault tolerance at small code distances using neural network-based decoders, demonstrating that the neural network can generalize to inputs that were not provided during training and that they can reach similar or better decoding performance compared to previous algorithms.
We conclude by discussing the time required by a feedforward neural network decoder in hardware.
\end{abstract}
\maketitle

\section{Introduction}
Quantum computing has emerged as a solution to accelerate various calculations using systems governed by quantum mechanics.
Such calculations are believed to take exponential time to perform using classical computers.
Initial applications where quantum computing will be useful are simulation of quantum physics \cite{simulation_review}, cryptanalysis \cite{shor, shor_factoring} and unstructured search \cite{grover}, and there is a growing set of other quantum algorithms \cite{quantum_zoo}.

Simple quantum algorithms have been shown to scale better than classical algorithms \cite{deutsch_jozsa, small_grover, quantum_simulation} for small test cases, though larger computers are required to solve real-world problems.
The main obstacle to scalability is that the required quantum operations (state preparations, single- and two-qubit unitary gates, and measurements) are subject to external noise, therefore quantum algorithms cannot run with perfect fidelity.
This requires quantum computers to use active error correction \cite{early_qec} to achieve scalability, which in turn requires a classical co-processor to infer which corrections to make, given a stream of measurement results as input. 
If this co-processor is slow, performance of the quantum computer may be degraded (though recent results \cite{chamberland} suggest that this may be mitigated).

The remainder of this paper is organized as follows.
In Section \ref{sec:QuantumErrorCorrection}, we outline the relevant aspects of quantum error correction and fault tolerance. 
We discuss the need for a fast classical co-processor in Section \ref{sec:NeedForFastDecoding}. 
In Section \ref{sec:RelatedWork}, we give a brief summary of existing techniques to perform decoding quickly, and follow this in Section \ref{sec:NeuralNetworkDecoder} with the introduction of a new technique based on feedforward neural networks.
We examine the accuracy of the proposed decoder in Section \ref{sec:Results}, and conclude by discussing its speed in Section \ref{sec:DiscussionAndConclusion}.

\section{Quantum error correction}
\label{sec:QuantumErrorCorrection}
While it is often possible to decrease the amount of noise affecting a quantum operation using advanced control techniques \cite{refocusing, Wilhelm}, their analog nature suggests that some imperfection will always remain. 
This has driven the development of algorithmic techniques to protect quantum states and computations from noise, which are called \emph{quantum error correction} and \emph{fault tolerance}, respectively. 

Quantum error correction replaces unprotected qubit states (e.g. $\ket{0}$, $\ket{1}$) with specially encoded multi-qubit entangled states (typically called $\ket{\bar{0}}$, $\ket{\bar{1}}$), such that a random operation $E$ acting on fewer than $d$ qubits cannot transform one encoded state into another ($\bra{\bar{0}}E \ket{\bar{1}} = 0$), where $d$ is called the \emph{code distance} \cite{gottesman, lidar}.
Typically, these random operations are taken to be \emph{Pauli operators}, whose names and effects on single-qubit states are given below:
\begin{equation}
X: \begin{array}{c}
\ket{0} \mapsto \ket{1} \\
\ket{1} \mapsto \ket{0}
\end{array}, \,\,
Z: \begin{array}{c}
\ket{0} \mapsto \ket{0} \\
\ket{1} \mapsto -\ket{1}
\end{array}, \,\,
Y: \begin{array}{c}
\ket{0} \mapsto i\ket{1} \\
\ket{1} \mapsto -i\ket{0}
\end{array}
\end{equation}
These operators form a convenient basis for the space of possible errors; codes which can correct these errors on a subset of qubits can correct arbitrary errors on the same subset \cite[Chapter 2]{lidar}.

Often, the encoded states are chosen to be in the mutual $+1$ eigenspace of a set of multi-qubit Pauli operators, called \emph{stabilisers}, resulting in a \emph{stabiliser code}.
Projective measurements of the stabilisers are used by a reliable classical co-processor to determine which error has occurred, a process called \emph{decoding}. 
The use of stabiliser codes can effectively reduce the probability of error from transmitting a qubit through a noisy channel (though \emph{logical errors} can still occur, acting as $\bar{X}$,  $\bar{Z}$, or $\bar{Y}$).
This reduction in the probability of error is obtained when operations are perfect, however, quantum error correction is not enough on its own to guarantee that computation can be performed with a low probability of error when using noisy operations.

To suppress errors from the physical operations themselves, it is necessary to design logical operations which act directly on encoded states (i.e. without first transforming the encoded states to bare qubit states), in such a way that random errors affecting physical operations are likely to result in correctable errors with respect to the underlying code. 
Operations which have this property are called fault-tolerant.
Fault-tolerant syndrome measurements can be applied repeatedly to correct time-dependent errors, which occur continuously as computation proceeds.
There are many schemes for attaining fault tolerance, based on different families of quantum codes, and using different techniques for ensuring noise from imperfect state preparation and stabiliser measurement remains suppressable.

Each fault tolerance scheme has a \emph{threshold error rate}, beneath which there exists a code in the associated code family which can suppress errors to exponential accuracy, using a polynomially-large number of qubits and operations \cite{threshold}.
Each code in such a family also typically has a \emph{pseudo-threshold}, an error rate at which encoded operations using that specific code provide higher accuracy than is possible using bare qubits/operations.
These figures of merit are used to characterize fault tolerance schemes, and are especially important when considering near-term implementations of these schemes.

\begin{figure}
\centering
\raisebox{-0.5\height}{\scalebox{0.8}{\begin{tikzpicture}[x = 12pt, y = 12pt, gridline/.style = {black, line width = 2pt, line join = round, line cap = round}]
\begin{scope}[shift = {(0, 0)}]
\foreach \x/\y in {1/1, 3/3}{
    \fill[white] (\x, \y) rectangle +(2,2);
}
\foreach \x/\y in {1/3, 3/1}{
    \fill[black!40!white] (\x, \y) rectangle +(2,2);
}
\draw[gridline, step=2, shift = {(-1,-1)}] (2, 2) grid (6, 6);
\foreach \x/\y in {3/5}{
    \filldraw[gridline, fill = white] (\x, \y) arc (0:180:1) -- cycle;
}
\foreach \x/\y in {5/3}{
    \filldraw[gridline, fill = black!40!white] (\x, \y) arc (-90:90:1) -- cycle;
}
\foreach \x/\y in {1/3}{
    \filldraw[gridline, fill = black!40!white] (\x, \y) arc (90:270:1) -- cycle;
}
\foreach \x/\y in {3/1}{
    \filldraw[gridline, fill = white] (\x, \y) arc (180:360:1) -- cycle;
}
\foreach \x in {1, 3, 5}{
    \foreach \y in {1, 3, 5}{
        \fill[black] (\x, \y) circle (3 pt) {}; 
    }
}
\end{scope}
\end{tikzpicture}}} \hspace{0.05\textwidth}
\raisebox{-0.5\height}{\scalebox{0.8}{\begin{tikzpicture}[x = 12pt, y = 12pt, gridline/.style = {black, line width = 2pt, line join = round, line cap = round}]
\begin{scope}[shift = {(0, 0)}]
\foreach \x/\y in {1/1, 1/5, 3/3, 3/7, 5/1, 5/5, 7/3, 7/7}{
    \fill[white] (\x, \y) rectangle +(2,2);
}
\foreach \x/\y in {1/3, 1/7, 3/1, 3/5, 5/3, 5/7, 7/1, 7/5}{
    \fill[black!40!white] (\x, \y) rectangle +(2,2);
}
\draw[gridline, step=2, shift = {(-1,-1)}] (2, 2) grid (10, 10);
\foreach \x/\y in {3/9, 7/9}{
    \filldraw[gridline, fill = white] (\x, \y) arc (0:180:1) -- cycle;
}
\foreach \x/\y in {9/3, 9/7}{
    \filldraw[gridline, fill = black!40!white] (\x, \y) arc (-90:90:1) -- cycle;
}
\foreach \x/\y in {1/3, 1/7}{
    \filldraw[gridline, fill = black!40!white] (\x, \y) arc (90:270:1) -- cycle;
}
\foreach \x/\y in {3/1, 7/1}{
    \filldraw[gridline, fill = white] (\x, \y) arc (180:360:1) -- cycle;
}
\foreach \x in {1, 3, 5, 7, 9}{
    \foreach \y in {1, 3, 5, 7, 9}{
        \fill[black] (\x, \y) circle (3 pt) {}; 
    }
}
\end{scope}
\end{tikzpicture}}} \hspace{0.05\textwidth}
\raisebox{-0.5\height}{\scalebox{0.8}{\begin{tikzpicture}[x = 12pt, y = 12pt, gridline/.style = {black, line width = 2pt, line join = round, line cap = round}]
\begin{scope}[shift = {(0, 0)}]
\foreach \x/\y in {1/1, 1/5, 1/9, 3/3, 3/7, 3/11, 5/1, 5/5, 5/9, 7/3, 7/7, 7/11, 9/1, 9/5, 9/9, 11/3, 11/7, 11/11}{
    \fill[white] (\x, \y) rectangle +(2,2);
}
\foreach \x/\y in {1/3, 1/7, 1/11, 3/1, 3/5, 3/9, 5/3, 5/7, 5/11, 7/1, 7/5, 7/9, 9/3, 9/7, 9/11, 11/1, 11/5, 11/9}{
    \fill[black!40!white] (\x, \y) rectangle +(2,2);
}
\draw[gridline, step=2, shift = {(-1,-1)}] (2, 2) grid (14, 14);
\foreach \x/\y in {3/13, 7/13, 11/13}{
    \filldraw[gridline, fill = white] (\x, \y) arc (0:180:1) -- cycle;
}
\foreach \x/\y in {13/3, 13/7, 13/11}{
    \filldraw[gridline, fill = black!40!white] (\x, \y) arc (-90:90:1) -- cycle;
}
\foreach \x/\y in {1/3, 1/7, 1/11}{
    \filldraw[gridline, fill = black!40!white] (\x, \y) arc (90:270:1) -- cycle;
}
\foreach \x/\y in {3/1, 7/1, 11/1}{
    \filldraw[gridline, fill = white] (\x, \y) arc (180:360:1) -- cycle;
}
\foreach \x in {1, 3, 5, 7, 9, 11, 13}{
    \foreach \y in {1, 3, 5, 7, 9, 11, 13}{
        \fill[black] (\x, \y) circle (3 pt) {}; 
    }
}
\end{scope}
\end{tikzpicture}}} 
\caption{Surface codes with distances $3$, $5$, and $7$, respectively.
Data qubits are placed at the corners of the square tiles, on which the stabilisers are supported.
White and grey squares support stabilisers of the form $X^{\otimes 4}$ and $Z^{\otimes 4}$, respectively. White and grey semi-circles support stabilisers of the form $X^{\otimes 2}$ and $Z^{\otimes 2}$, respectively.
Ancilla qubits placed inside the tiles can be coupled to neighbouring data qubits and measured to effect indirect stabiliser measurement. }
\label{fig:SCs}
\end{figure}
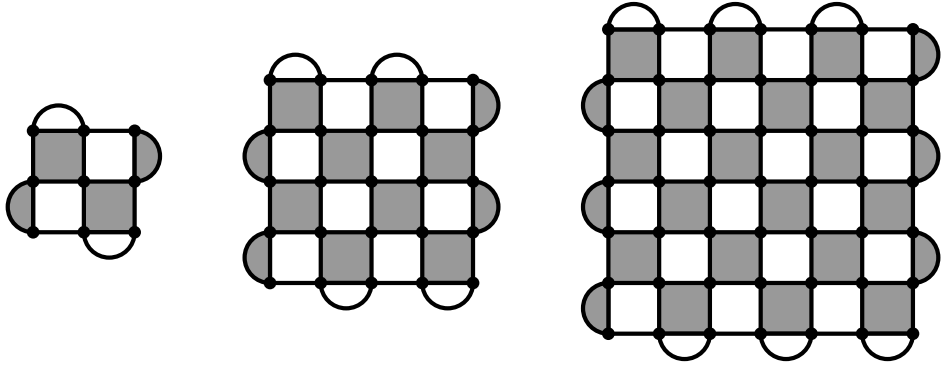

\begin{figure}
\centering
\begin{minipage}{0.3\textwidth}
\begin{tikzpicture}[scale=1.500000,x=1pt,y=1pt]
\filldraw[color=white] (0.000000, -4.500000) rectangle (66.000000, 40.500000);
\draw[color=black] (0.000000,36.000000) -- (66.000000,36.000000);
\draw[color=black] (0.000000,27.000000) -- (66.000000,27.000000);
\draw[color=black] (0.000000,18.000000) -- (66.000000,18.000000);
\draw[color=black] (0.000000,9.000000) -- (66.000000,9.000000);
\draw[color=black] (0.000000,0.000000) -- (57.000000,0.000000);
\draw[color=black] (57.000000,-0.500000) -- (66.000000,-0.500000);
\draw[color=black] (57.000000,0.500000) -- (66.000000,0.500000);
\draw[color=black] (0.000000,0.000000) node[left] {$\ket{0}$};
\draw (6.000000,36.000000) -- (6.000000,0.000000);
\begin{scope}
\draw[fill=white] (6.000000, 0.000000) circle(3.000000pt);
\clip (6.000000, 0.000000) circle(3.000000pt);
\draw (3.000000, 0.000000) -- (9.000000, 0.000000);
\draw (6.000000, -3.000000) -- (6.000000, 3.000000);
\end{scope}
\filldraw (6.000000, 36.000000) circle(1.500000pt);
\draw (18.000000,27.000000) -- (18.000000,0.000000);
\begin{scope}
\draw[fill=white] (18.000000, 0.000000) circle(3.000000pt);
\clip (18.000000, 0.000000) circle(3.000000pt);
\draw (15.000000, 0.000000) -- (21.000000, 0.000000);
\draw (18.000000, -3.000000) -- (18.000000, 3.000000);
\end{scope}
\filldraw (18.000000, 27.000000) circle(1.500000pt);
\draw (30.000000,18.000000) -- (30.000000,0.000000);
\begin{scope}
\draw[fill=white] (30.000000, 0.000000) circle(3.000000pt);
\clip (30.000000, 0.000000) circle(3.000000pt);
\draw (27.000000, 0.000000) -- (33.000000, 0.000000);
\draw (30.000000, -3.000000) -- (30.000000, 3.000000);
\end{scope}
\filldraw (30.000000, 18.000000) circle(1.500000pt);
\draw (42.000000,9.000000) -- (42.000000,0.000000);
\begin{scope}
\draw[fill=white] (42.000000, 0.000000) circle(3.000000pt);
\clip (42.000000, 0.000000) circle(3.000000pt);
\draw (39.000000, 0.000000) -- (45.000000, 0.000000);
\draw (42.000000, -3.000000) -- (42.000000, 3.000000);
\end{scope}
\filldraw (42.000000, 9.000000) circle(1.500000pt);
\draw[fill=white] (51.000000, -4.000000) -- (59.000000,-4.000000) arc (-90:90:4.000000pt) -- (51.000000,4.000000) -- cycle;
\draw (57.000000, 0.000000) node {{\small $Z$}};
\end{tikzpicture}

\begin{tikzpicture}[scale=1.500000,x=1pt,y=1pt]
\filldraw[color=white] (0.000000, -4.500000) rectangle (66.000000, 40.500000);
\draw[color=black] (0.000000,36.000000) -- (66.000000,36.000000);
\draw[color=black] (0.000000,27.000000) -- (66.000000,27.000000);
\draw[color=black] (0.000000,18.000000) -- (66.000000,18.000000);
\draw[color=black] (0.000000,9.000000) -- (66.000000,9.000000);
\draw[color=black] (0.000000,0.000000) -- (57.000000,0.000000);
\draw[color=black] (57.000000,-0.500000) -- (66.000000,-0.500000);
\draw[color=black] (57.000000,0.500000) -- (66.000000,0.500000);
\draw[color=black] (0.000000,0.000000) node[left] {$\ket{+}$};
\draw (6.000000,36.000000) -- (6.000000,0.000000);
\begin{scope}
\draw[fill=white] (6.000000, 36.000000) circle(3.000000pt);
\clip (6.000000, 36.000000) circle(3.000000pt);
\draw (3.000000, 36.000000) -- (9.000000, 36.000000);
\draw (6.000000, 33.000000) -- (6.000000, 39.000000);
\end{scope}
\filldraw (6.000000, 0.000000) circle(1.500000pt);
\draw (18.000000,27.000000) -- (18.000000,0.000000);
\begin{scope}
\draw[fill=white] (18.000000, 27.000000) circle(3.000000pt);
\clip (18.000000, 27.000000) circle(3.000000pt);
\draw (15.000000, 27.000000) -- (21.000000, 27.000000);
\draw (18.000000, 24.000000) -- (18.000000, 30.000000);
\end{scope}
\filldraw (18.000000, 0.000000) circle(1.500000pt);
\draw (30.000000,18.000000) -- (30.000000,0.000000);
\begin{scope}
\draw[fill=white] (30.000000, 18.000000) circle(3.000000pt);
\clip (30.000000, 18.000000) circle(3.000000pt);
\draw (27.000000, 18.000000) -- (33.000000, 18.000000);
\draw (30.000000, 15.000000) -- (30.000000, 21.000000);
\end{scope}
\filldraw (30.000000, 0.000000) circle(1.500000pt);
\draw (42.000000,9.000000) -- (42.000000,0.000000);
\begin{scope}
\draw[fill=white] (42.000000, 9.000000) circle(3.000000pt);
\clip (42.000000, 9.000000) circle(3.000000pt);
\draw (39.000000, 9.000000) -- (45.000000, 9.000000);
\draw (42.000000, 6.000000) -- (42.000000, 12.000000);
\end{scope}
\filldraw (42.000000, 0.000000) circle(1.500000pt);
\draw[fill=white] (51.000000, -4.000000) -- (59.000000,-4.000000) arc (-90:90:4.000000pt) -- (51.000000,4.000000) -- cycle;
\draw (57.000000, 0.000000) node {{\small $X$}};
\end{tikzpicture}
\end{minipage}
\begin{minipage}{0.6\textwidth}
\includegraphics[width=\textwidth]{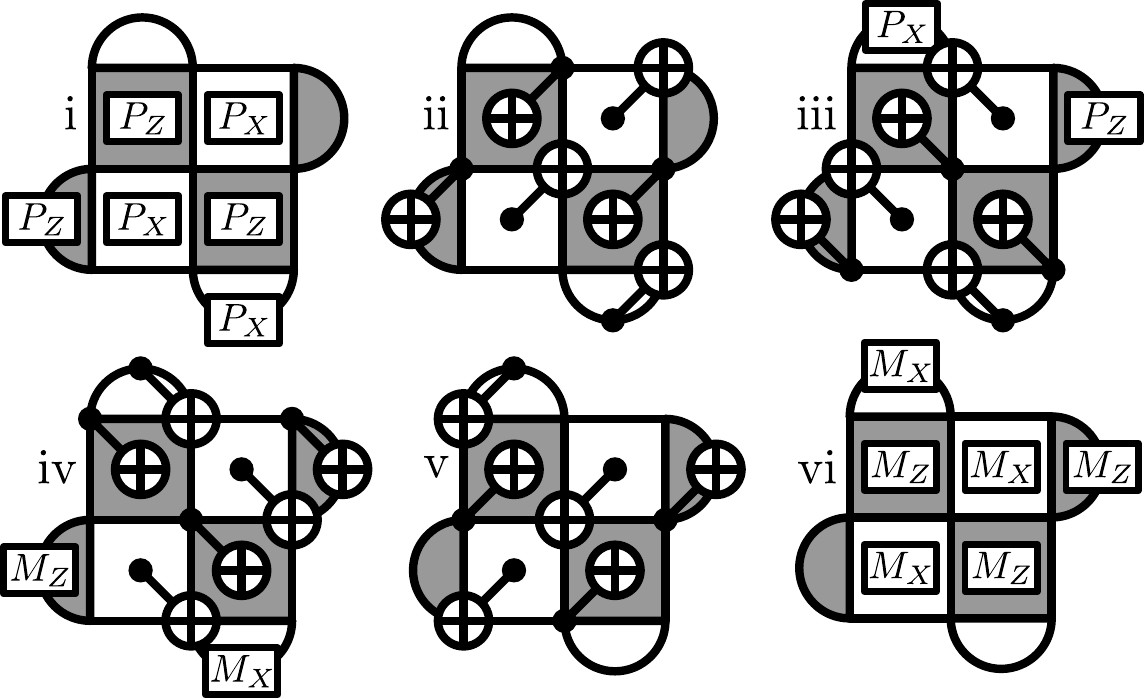}
\end{minipage}
\caption{Stabiliser measurement circuit for the distance-$3$ surface code \cite{dklp, fowler2009high, tomita}.
Left: Measurement circuit for individual $Z$ tiles (top) and $X$ tiles (bottom), including an ancilla qubit to be placed at the center of each tile. 
Ancilla qubits are prepared in the $+1$-eigenstate of the appropriate basis, four \textsc{cnot} gates are executed, and the ancilla qubits are measured in the appropriate basis.
Right: Interleaving of separate stabiliser measurements, including late preparation and early measurement for weight-two stabilisers.}
\label{fig:MeasCirc}
\end{figure}
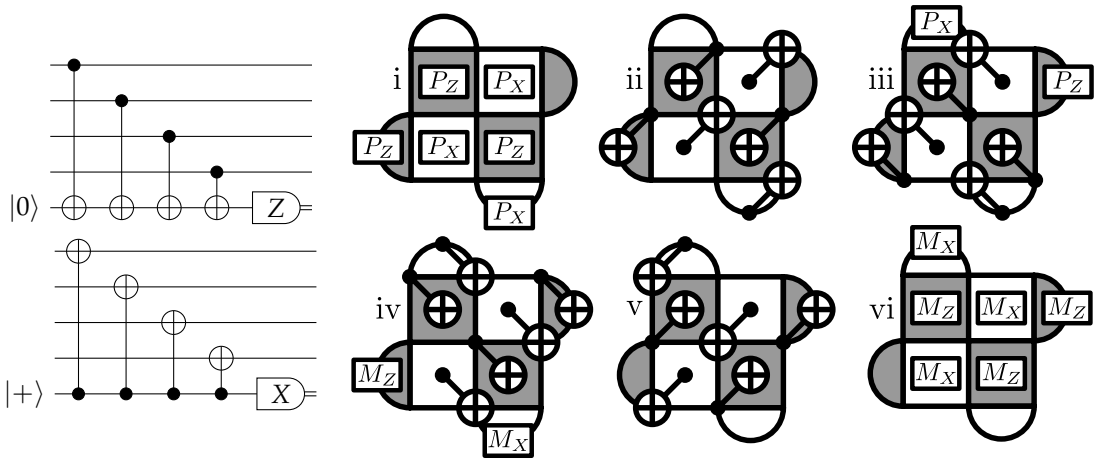

One scheme which has a relatively high threshold error rate uses \emph{surface codes} \cite{Kitaev2003, freedman2001projective, bravyi1998quantum, bombin2007optimal}, stabiliser codes whose stabilisers are supported on qubits which are adjacent on a 2D square tiling (see Figure \ref{fig:SCs}).
This approach also allows the use of exclusively planar connections between qubits, and uses at most four connections between each qubit and its neighbours (see Figure \ref{fig:MeasCirc}). These features make surface codes especially attractive for near-term implementation.

\begin{figure}
\centering
\includegraphics[width=0.375\textwidth]{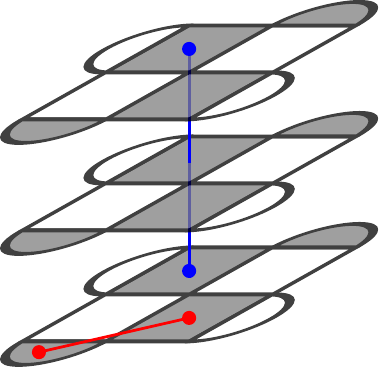}
\caption{Three consecutive rounds of surface code measurement arranged in a $2+1$-dimensional lattice.
Errors on data qubits result in horizontally-separated changes in the syndrome record, measurement errors result in vertically-separated changes.}
\label{fig:Syndromes}
\end{figure}

To complete such an implementation and analyse its performance, it is also necessary to specify the method by which surface codes are to be decoded. 
Syndromes obtained by measuring surface code stabilisers have a special mathematical structure, which leads to a polynomial-time decoding algorithm. 
These syndromes occur at the endpoints of continuous one-dimensional chains of errors if stabiliser measurement is performed with perfect operations, and differences between consecutive syndromes occur at the endpoints of one-dimensional chains of data/measurement errors if realistically noisy operations are used (see Figure \ref{fig:Syndromes}).
If error rates are low, then the smallest error which conforms with the syndrome is likely a valid correction. 
To find it requires the classical co-processor to minimize the sum of the lengths of chains connecting pairs of syndrome changes, a problem known as \emph{minimum-weight perfect matching} \cite{schrijver2003combinatorial}. 
This problem can be solved using the \emph{Blossom algorithm} \cite{edmonds, kolmogorov}. 
This algorithm produces accurate corrections for the surface code, but has a complexity which scales polynomially with respect to $d$. 
This is an obstacle to using the Blossom algorithm for decoding surface codes in practice, for reasons which we explain in the following section.

\section{Need for fast decoding}
\label{sec:NeedForFastDecoding}
Projective measurement of the logical qubits and classical feedforward of the measurement values are key ingredients in universal fault-tolerant quantum computing. To calculate the bit which we feed forward, we need to decode. Thus, it is necessary to decode frequently during a computation.

While the decoding takes place at the classical co-processor, we could either continue running rounds of syndrome measurement or stop and wait for the decoding to be concluded. If we stop the computation, errors will build up until they become uncorrectable. This takes an amount of time which depends on the implementation in question ($\sim 10 \,\mu\textrm{s}$ in current superconducting circuits, for example \cite{toms_fancy_simulation_paper}).
On the other hand, if we continue measuring syndromes, we will build a \emph{backlog} of data that produces a more difficult decoding problem in the future.
The ideal case would be a decoder that decodes $d$ rounds of syndrome measurement in less time than the time needed to perform the measurements themselves. In superconducting circuits, the time for a single round of syndrome measurement is $800 \textrm{ ns}$ \cite{leos_fancy_cycle_paper}.

There are many techniques that provide high performance decoding.
In the following section, we summarize some of them.

\section{Related work}
\label{sec:RelatedWork}
To decrease decoding time when correcting time-dependent errors, the ``overlapping recovery'' method was introduced in \cite{dklp}.
This method divides the measurement record into \emph{windows}, defined as a set of $\sim d$ consecutive error correction cycles.
In the overlapping recovery technique, syndromes are matched either to each other (pairwise) or to a time boundary placed immediately after the last round of syndrome measurement.
At the next window, the syndromes matched to the time boundary are forwarded to the following window, in order to identify chains of errors which cross the boundary.
This reduces the backlog problem mentioned earlier, by allowing the decoding problem to be solved incrementally.

To further reduce the backlog, Fowler \cite{fowler_o1} has parallelized the Blossom algorithm, using message-passing between local processors to replace slow subroutines.
This technique produces accurate corrections, resulting in a high threshold error rate, and is scalable to large code distances. 
However, in the near future, only small code distances will be experimentally viable, so it is likely that a heuristic approach will perform well. 

One such approach is taken in \cite{tomita}.
In this paper, the authors have designed a heuristic-based decoder that resembles the parallelized MWPM decoding for a distance-3 Surface Code with a window of 3 error correction cycles. 
The simple structure of this heuristic algorithm makes it easily programmable to hardware, decreasing the decoding time.
The main drawback of this algorithm is that it cannot easily be extended to higher code distances, so an alternate method is required.

Currently, machine learning techniques are being explored as possible alternate decoding techniques, independently of the need for high-speed decoding.
One such technique is being used in \cite{torlai}.
The authors of this paper use a stochastic neural network (or Boltzmann machine) to decode stabilizer codes.
They optimize the neural network to fit a dataset that includes the errors and their respective syndromes.
The network then models the probability distribution of the errors in the dataset and generates prospective recovery error chains when a syndrome is input.
Many networks are produced for a variety of physical error probabilities $p$, so when an error syndrome is obtained, a random recovery chain of errors is sampled from the distribution corresponding to the known value of $p$.
While this method produced similar performance to MWPM decoding for simple error models, repeated sampling is required in order to produce an error that conforms with the syndrome, which takes unknown time.

To achieve high performance in bounded time, we use a simpler machine learning technique, the feed-forward neural network, which we introduce and apply to the decoding problem in the next section.

\section{Neural network decoder}
\label{sec:NeuralNetworkDecoder}
To apply machine learning techniques to surface code decoding, we first reduce the decoding problem to a well-studied problem in machine learning; classification. 
Classification problems consist of a set of (generally high-dimensional) inputs, each of which is associated with a (generally low-dimensional) label. 
The goal is to optimize the assignment of known labels to known inputs (a process called \emph{training}) so that unknown inputs can also be correctly labeled. 

To reduce the decoding problem to a classification problem, we decompose an error $E$ into three multi-qubit Pauli operators:
\begin{equation}
E = S \cdot C \cdot L,
\end{equation}
where $S$ is a stabiliser, $C$ is any unique Pauli which produces the syndrome $\vec{s}$ (also known as a \emph{pure error} \cite{pure_errors}), and $L$ is a logical Pauli operator of the surface code.  
Any decoder which provides a correction $E' = S' \cdot C \cdot L$, in which the stabiliser in the correction is different from that in the actual error, does not lead to a logical error.
This implies that $S$ can be assigned arbitrarily with no impact on decoder performance. 
Also, it is possible to produce a pure error by parallel table look-up, since each bit of the syndrome can be assigned a unique pure error, independently of the other bits. 
We call the apparatus that produces this error the \emph{simple decoder}.
Since pure errors can be determined quickly in this fashion, the neural network only has to identify $L$, which can take one of four values; $\id$, $\bar{X}$, $\bar{Y}$, or $\bar{Z}$. 
These four values can be used as labels in a classification problem.

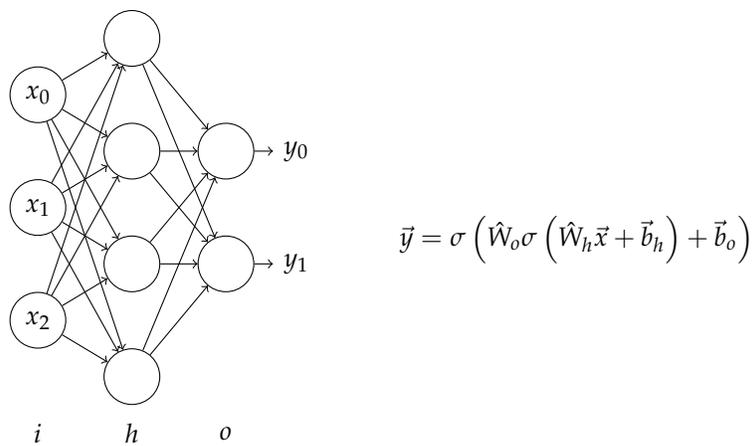
\begin{figure}[!h]
\centering
\raisebox{-0.5\height}{\begin{tikzpicture}[yscale=0.75, xscale=1.25]
\foreach \y/\lbl in {1/2, 3/1, 5/0}{
    \node (n_0_\y) at (0, \y) [circle, draw=black] {$x_\lbl$};
}
\foreach \y in {0, 2, 4, 6}{
    \node (n_1_\y) at (1, \y) [circle, draw=black] {\phantom{$x_0$}};
    \foreach \yp in {1, 3, 5}{
        \draw [black, ->] (n_0_\yp) -- (n_1_\y);
    }
}
\foreach \y in {2, 4}{
    \node (n_2_\y) at (2, \y) [circle, draw=black] {\phantom{$x_0$}};
    \foreach \yp in {0, 2, 4, 6}{
        \draw [black, ->] (n_1_\yp) -- (n_2_\y);
    }
    \draw [black, ->] (n_2_\y) -- (2.5, \y);
}
\node at (0, -1) {$i$};
\node at (1, -1) {$h$};
\node at (2, -1) {$o$};
\node at (2.75, 2) {$y_1$};
\node at (2.75, 4) {$y_0$};
\end{tikzpicture}} \hspace{0.05\textwidth}
\raisebox{-0.5\height}{$\vec{y} = \sigma \left( \hat{W}_o \sigma \left( \hat{W}_h \vec{x} + \vec{b}_h \right) + \vec{b}_o \right)$}
\caption{The graphical and functional descriptions of a feed-forward neural network.
In the graphical description (left), inputs $x_j$ are passed to neurons in a \emph{hidden layer}, and each of these neurons outputs $\sigma \left( \vec{w}\cdot \vec{x} + b \right)$, where $\vec{w}$  and $b$ are a local set of weights and a bias, and $\sigma(x)$ is a non-linear \emph{activation function} (we use $\sigma(x) = (1 + \exp(x))^{-1}$ for all neurons considered in this work).
The final outputs $y_k$ can be rounded to $\left \lbrace 0, \, 1 \right \rbrace$, and interpreted as a class label.
In the functional picture, the weights and biases are assembled into matrices and vectors, respectively, allowing the output vector to be expressed as a composition of functions acting on the input vector. 
}
\label{fig:NNschematic}
\end{figure}

To solve this problem, we use feed-forward neural networks, which are widely regarded as the simplest machine learning technique \cite{ml_textbook}.
A feed-forward neural net can be described graphically or functionally, see Figure \ref{fig:NNschematic}.
In either description, a feed-forward neural network contains a large number of free parameters, which are assigned values which minimize a given cost function. 
A typical cost function, which we use in this work, is the average cross-entropy:
\begin{equation}
\left \langle H(p, y) \right \rangle \propto - \sum_{(\vec{p}, \vec{x}) \in T} \vec{p} \cdot \ln(\vec{y}(\vec{x})),
\end{equation}
where $T$ is the training set, consisting of desired (`target') distributions $\vec{p}$ and input values $\vec{x}$. 
To minimize this function, we use stochastic gradient descent, as implemented in the Tensorflow library \cite{tensorflow}.
To produce a training set, we use direct sampling at a single physical error probability, where the Blossom algorithm produces a logical error rate of $\sim 25\%$. This physical error probability is chosen so that a large variety of error syndromes can be produced while still ensuring that correction is possible.
For small surface codes, it is possible to sample the entire set of possible syndromes, we limit the size of the training set to at most $10^6$ samples for larger codes. This training set size provides relatively fast training and high accuracy, as seen in Section \ref{sec:Results}.

In the following section, we compare the performance of our decoder to the performance of Blossom and the performance of the partial lookup table (PLUT), which contains the error syndromes and corrections from the training set, returning the most likely correction ($\id$) for error syndromes that are not in the training set. The comparison in terms of performance is based on the logical error rate of each decoder for specific code distances and error models.

\section{Results}
\label{sec:Results}
In the proposed decoder, we provide the error syndrome to both the simple decoder and the neural network. 
As presented in Table 1, the size of the input for the neural network is equal to the number of required syndrome bits, depending on the error model, and only one hidden layer was deemed adequate for all networks.
The number of nodes in the hidden layer was decided based on the performance of the neural network during training and testing.
\begin{table}[!h]
\centering
\begin{tabular}{ |c|c|c|c|  }\hline
\multicolumn{4}{|c|}{QEC Error Models} \\\hline
\multicolumn{4}{|c|}{Channel Capacity Noise} \\\hline
Code distance & Input nodes & Hidden nodes & Output nodes \\\hline
3 & 4 & 10 & 2\\ \hline
5 & 12 & 90 & 2\\ \hline
7 & 24 & 512 & 2\\ \hline
\multicolumn{4}{|c|}{Depolarizing Noise} \\\hline
3 & 8 & 128 & 4\\ \hline
5 & 24 & 660 & 4\\ \hline
7 & 48 & 256 & 4\\ \hline
\end{tabular}

\begin{tabular}{ |c|c|c|c|  }\hline
\multicolumn{4}{|c|}{FT Error Models} \\\hline
\multicolumn{4}{|c|}{Channel Capacity \& Measurement Noise} \\\hline
Code distance & Input nodes & Hidden nodes & Output nodes \\\hline
3 & 16 & 768 & 4\\ \hline
\multicolumn{4}{|c|}{Depolarizing \& Measurement Noise} \\\hline
3 & 32 & 768 & 4\\ \hline
\multicolumn{4}{|c|}{Circuit noise} \\\hline
3 & 32 & 704 & 4\\ \hline
\end{tabular}
\caption{Layer sizes for the neural networks used throughout this work.
The number of input nodes is determined by the number of syndromes in the quantum error correction scenario, using only $X$ (or $Z$) syndrome bits for independent $X$/$Z$ errors, and all syndrome bits for depolarizing errors. 
In the fault tolerance scenario, $d$ rounds of measurement are followed by readout of the data qubits, and calculated stabiliser eigenvalues are included in the input.
The output layer is restricted to two nodes for independent $X$/$Z$ errors, since logical $X$/$Z$ errors are also independent. 
In all other scenarios, four nodes are used to discriminate between $\id$, $\bar{X}$, $\bar{Y}$, and $\bar{Z}$.
The number of nodes in the hidden layer is determined by analysing the performance of the resulting decoder empirically.}
\end{table}

We test the proposed decoder against Blossom and the PLUT decoder for two classes of error models, called quantum error correction (QEC) and fault tolerance (FT). 
Quantum error correction (QEC) error models approximate noise only on data qubits and fault tolerance (FT) error models approximate noise on gates and operations, therefore requiring multiple rounds of measurement to find all errors.
The channel capacity model inserts only $X$ or only $Z$ errors with probability $p$ in the data qubits.
The depolarizing model places $X$/$Y$/$Z$ errors with equal probability, $\nicefrac{p}{3}$, on the data qubits. For these error models only one cycle of error correction is required to find all errors.

In the fault tolerance scenario, the probability of an error occurring on a qubit and the probability of a measurement error is the same, therefore the minimum number of rounds of measurement is taken to be $d$.
Instead of data qubit and measurement errors, the circuit noise model assumes that all operations and gates are noisy.
Each single-qubit gate is followed by depolarizing noise with probability $\nicefrac{p}{3}$ and each two-qubit gate is followed by a two-bit depolarizing map where each non-$\id \otimes \id$ two-bit Pauli has probability $\nicefrac{p}{15}$.
Preparation and measurement locations fail with probability $p$, resulting in a prepared $-1$-eigenstate or measurement error, respectively.

\begin{figure}[!h]
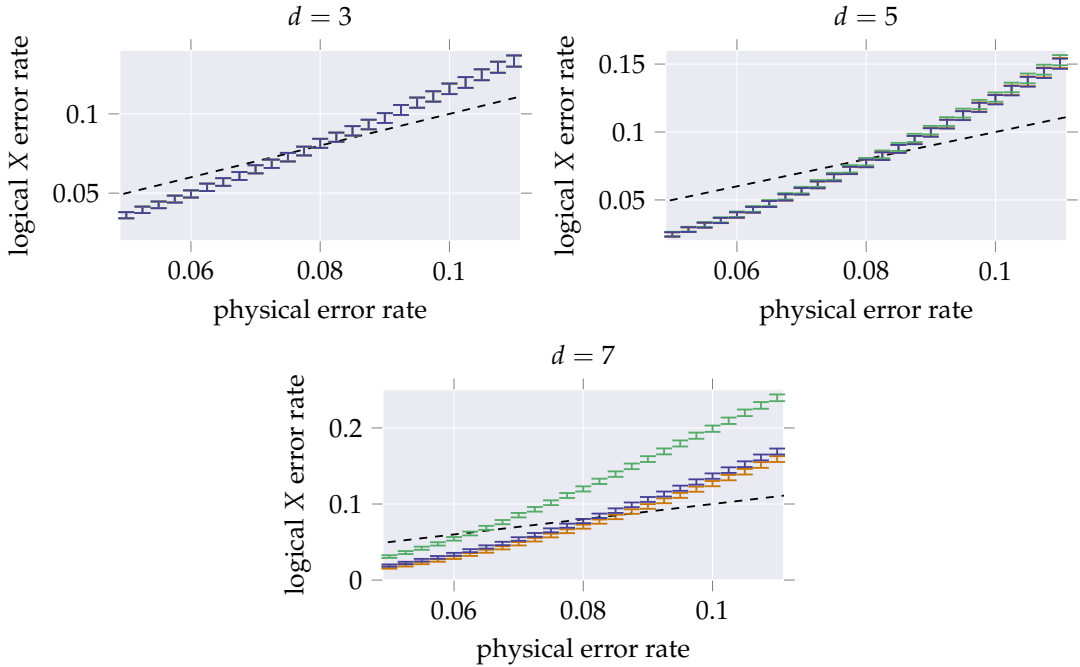

\centering 
\input{iid_3.tex}
\input{iid_5.tex}
\input{iid_7.tex}
\caption{Channel capacity error model without measurement errors for Surface Code distances 3,5 and 7. Performance comparison of the neural network decoder (blue) to the MWPM algorithm (orange) and partial look-up table (green).}\label{fig:iid_2d}
\end{figure}

\begin{figure}[!h]
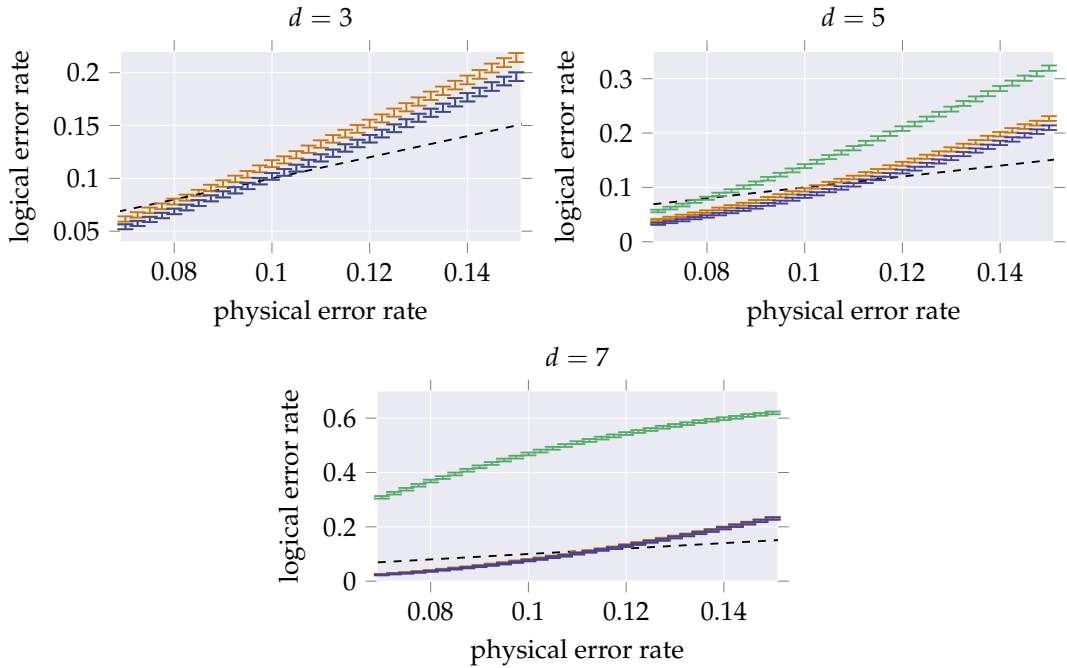

\centering 
\input{depol_3.tex}
\input{depol_5.tex}
\input{depol_7.tex}
\caption{Depolarizing error model without measurement errors for Surface Code distances 3,5 and 7. Performance comparison of the neural network decoder (blue) to the MWPM algorithm (orange) and partial look-up table (green).}\label{fig:depol_2d}
\end{figure}

\begin{figure}[!h]
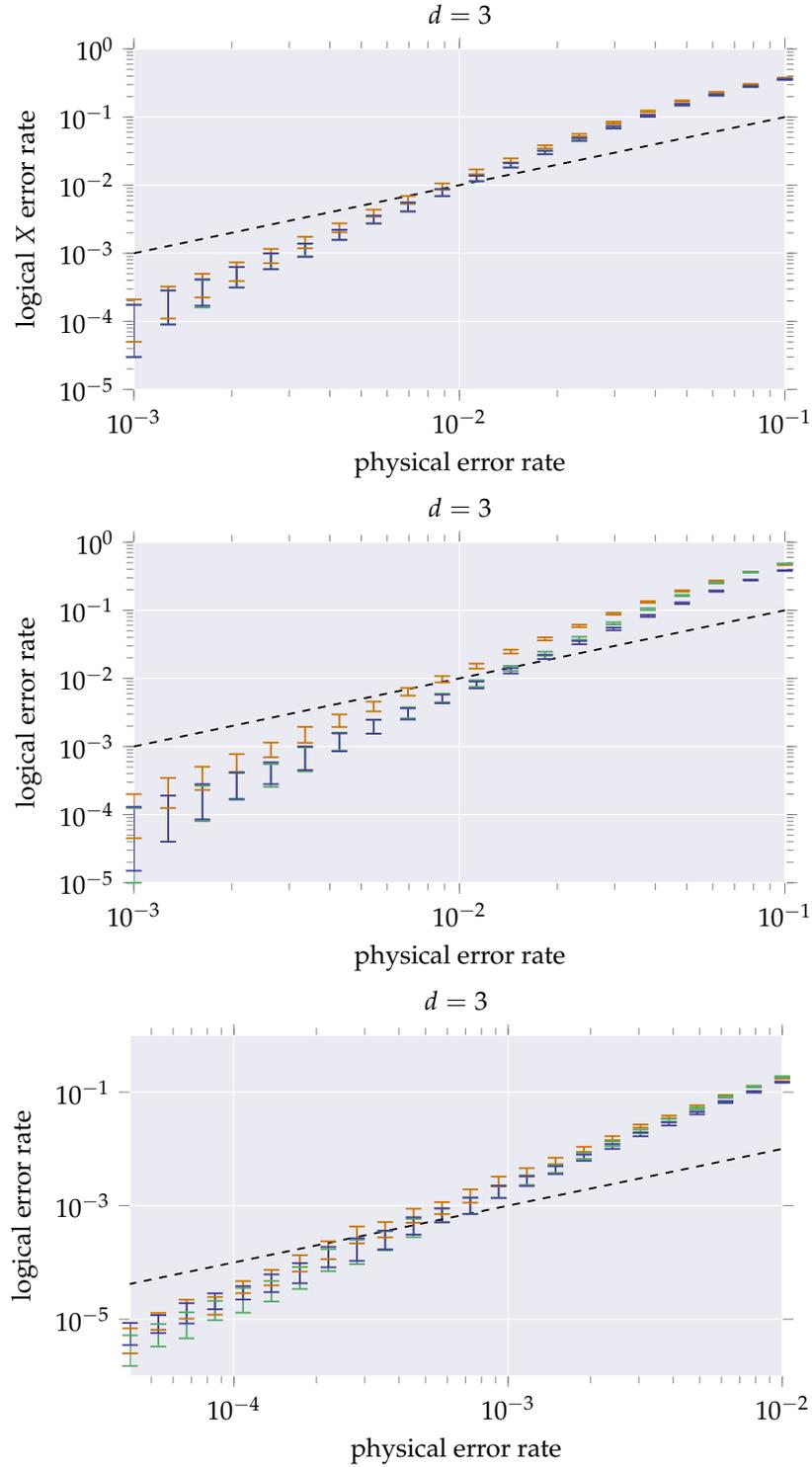

\centering 
\input{iid_3d_3.tex}
\input{dep_3d_3.tex}
\input{circ_3d_3.tex}
\caption{Channel capacity error model with measurement errors for Surface Code distance 3 (top). Depolarizing error model with measurement errors for Surface Code distance 3 (middle). Circuit error model for Surface Code distance 3 (bottom). Performance comparison of the neural network decoder (blue) to the MWPM algorithm (orange) and partial look-up table (green).}\label{fig:3D}
\end{figure}

In our simulations more than $10^6$ error correction cycles were run per point and each point has a confidence interval of 99.9\%. The percentage of the most frequent error syndromes that were used as training cases for the QEC error models were $100\%$ ($d=3$), $72.46\%$ ($d=5$), $2.75\%$ ($d=7$), see figure \ref{fig:iid_2d}, and $100\%$ ($d=3$), $0.98\%$ ($d=5$), $3 \times 10^{-7}\%$ ($d=7$), see figure \ref{fig:depol_2d}, for channel capacity and depolarizing models respectively. The percentage of the most frequent error syndromes that were used as training cases for the fault tolerance error models were $30.09\%$, $0.022\%$ and $0.01\%$ for the channel capacity (see figure \ref{fig:3D} top), the depolarizing (see figure \ref{fig:3D} middle), and the circuit model (see figure \ref{fig:3D} bottom), respectively.
The performance of our decoder was compared to the Blossom algorithm and the PLUT decoder.

In the QEC error models, see figure \ref{fig:iid_2d} and figure \ref{fig:depol_2d}, we observe a clear trend.
In both error models, as the distance increases the performance of our decoder remains similar to Blossom, and becomes much better that the PLUT-based decoder.
This demonstrates that the neural networks of our decoder can successfully correct error syndromes that were not included in training.
At small code distances, almost all possible error syndromes were used in training, resulting in identical performance from both the PLUT and our decoder.
However, going to larger distances while using a small set of error syndromes for training, leads to sub-optimal decoding by the PLUT decoder.

It is known that, for the channel capacity error model, Blossom can reach near-optimal performance, therefore it is sufficient for our decoder to reach similar performance. There are correctable errors (with weight $\leq 3$) in distance 7 that are not included in the training set and the neural network is not generalizing correctly. Therefore, the performance is slightly worse than Blossom's. However, for the depolarizing error model, Blossom is known to misidentify $Y$ errors, since it performs the decoding for $X$ and $Z$ errors separately, treating a $Y$ error as two distinct errors. Thus, if we train our decoder to take $Y$ errors into account as weight-1 errors, the performance will be better than Blossom's. In the depolarizing model, there are still a few weight 3 errors that are being mis-identified, however the existance of higher weight errors in the training set, that are being corrected properly, account for the sligthly better performance compared to the Blossom decoder.

In the fault tolerance scenario, see figure \ref{fig:3D}, due to the small code distance, all decoders reach a similar level of performance. Specifically, for the channel capacity and the depolarizing error model, a small amount of error syndromes was only necessary to reach Blossom's performance. The circuit metric required more syndromes, however slightly better performance was achieved in this case as well.

It is encouraging that the neural network based decoder can achieve similar performance to Blossom. However, the main reason that such a design is proposed is to accelerate the decoding time.
In the following section, we provide an estimation of the speed of the neural network based decoder in hardware, and discuss the implications for future research.

\section{Discussion and conclusion}
\label{sec:DiscussionAndConclusion}
In order to accurately estimate the execution time of the proposed decoder in hardware, we take advantage of the ability to perform many operations in parallel.
For example, in the simple decoder, the correction for each syndrome bit is independent of the other bits, allowing all of these corrections to be derived simultaneously. 
In order to determine whether to apply an $X$ to a given qubit, for example, it is then necessary to determine the parity of the number of corrections affecting the qubit. 
There are at most $\left \lfloor \nicefrac{d}{2} \right \rfloor$ possible corrections (this is the largest possible number of syndrome bits between an arbitrary syndrome bit and the boundary), and evaluating the parity of a set of $b$ bits requires a tree-like circuit of depth $\left \lceil \log_2(b) \right \rceil$ \texttt{xor} gates. 
In an FPGA (field-programmable gate array), simple operations such as \texttt{xor} can be evaluated in a single clock cycle, typically $2.5$--$5$ ns\cite{clock_cycle}. 
We can also take advantage of the graphical description of neural networks (see Figure \ref{fig:NNschematic}) to evaluate their output quickly. 
Firstly, the output of each neuron can be evaluated independently, so the runtime is dominated by the time needed to take an inner product between the weight vector and an intermediate state in the network. Each multiplication can be performed independently, and summation requires a logarithmic number of adders, similar to calculating parity. 
In a digital signal processing (DSP) slice, present in most FPGAs, simple arithmetic operations can also be carried out in a single clock cycle. 

For our largest neural network, then, we require two multiplications (one for the hidden layer, and one for the output layer), 15 addition steps ($\lceil \log_2(32) \rceil + \lceil \log_2(768) \rceil$), and two evaluations of the sigmoid function, for a total of $19$ serial steps. 
If each of these can be carried out in a single clock cycle, the time required will be $\sim 100$ ns.
This timing estimate is optimistic, since it does not account for the possibility that high-accuracy arithmetic may require additional clock cycles, or that communication between distant components of the FPGA may take longer than arithmetic operations.
However, if these factors increase the execution time by a factor of $\sim 20$, the proposed decoder will still be able to decode the syndrome from three consecutive rounds of measurement if each round requires $\sim 800$ ns.

Immediate future work will focus on implementing the proposed decoder in hardware, to determine the overhead caused by communication.
Once this overhead is reduced to an acceptable level, we can begin to extend the proposed decoder to the case where syndromes from a finite window are fed forward, as in \cite{dklp}.
In addition, we can begin testing the applicability of feedforward neural networks to surface codes with larger distance, as well as to alternate codes for which existing decoders do not attain high accuracy and speed simultaneously \cite{color_code_defn, color_code_ft, color_code_rg}. 

In conclusion, feedforward neural networks provide a fast and accurate method to decode small surface codes, both for performing quantum error correction, as well as fault-tolerant operations.
Given that the hardware requirements and anticipated runtime are relatively low, we expect feedfoward neural network decoders to be usable in the near term.

\bibliographystyle{unsrtnat}
\bibliography{arXiv_ref}

\begin{thebibliography}{36}
\providecommand{\natexlab}[1]{#1}
\providecommand{\url}[1]{\texttt{#1}}
\expandafter\ifx\csname urlstyle\endcsname\relax
  \providecommand{\doi}[1]{doi: #1}\else
  \providecommand{\doi}{doi: \begingroup \urlstyle{rm}\Url}\fi

\bibitem[Georgescu et~al.(2014)Georgescu, Ashhab, and Nori]{simulation_review}
IM~Georgescu, S~Ashhab, and Franco Nori.
\newblock Quantum simulation.
\newblock \emph{Reviews of Modern Physics}, 86\penalty0 (1):\penalty0 153,
  2014.

\bibitem[Shor(1997)]{shor}
Peter~W. Shor.
\newblock Polynomial-time algorithms for prime factorization and discrete
  logarithms on a quantum computer.
\newblock \emph{SIAM Journal on Computing}, 26\penalty0 (5):\penalty0
  1484--1509, 1997.
\newblock \doi{10.1137/S0097539795293172}.
\newblock URL \url{http://dx.doi.org/10.1137/S0097539795293172}.

\bibitem[Monz et~al.(2016)Monz, Nigg, Martinez, Brandl, Schindler, Rines, Wang,
  Chuang, and Blatt]{shor_factoring}
Thomas Monz, Daniel Nigg, Esteban~A. Martinez, Matthias~F. Brandl, Philipp
  Schindler, Richard Rines, Shannon~X. Wang, Isaac~L. Chuang, and Rainer Blatt.
\newblock Realization of a scalable shor algorithm.
\newblock \emph{Science}, 351\penalty0 (6277):\penalty0 1068--1070, 2016.
\newblock \doi{10.1126/science.aad9480}.

\bibitem[Grover(1997)]{grover}
Lov~K. Grover.
\newblock Quantum mechanics helps in searching for a needle in a haystack.
\newblock \emph{Phys. Rev. Lett.}, 79:\penalty0 325--328, Jul 1997.
\newblock \doi{10.1103/PhysRevLett.79.325}.
\newblock URL \url{https://link.aps.org/doi/10.1103/PhysRevLett.79.325}.

\bibitem[Jordan(2016)]{quantum_zoo}
Stephen Jordan.
\newblock Quantum algorithms {@ONLINE}, October 2016.
\newblock URL \url{http://math.nist.gov/quantum/zoo/}.

\bibitem[Wu et~al.(2011)Wu, Li, Zheng, Luo, Feng, and Peng]{deutsch_jozsa}
Zhen Wu, Jun Li, Wenqiang Zheng, Jun Luo, Mang Feng, and Xinhua Peng.
\newblock Experimental demonstration of the deutsch-jozsa algorithm in
  homonuclear multispin systems.
\newblock \emph{Phys. Rev. A}, 84:\penalty0 042312, Oct 2011.
\newblock \doi{10.1103/PhysRevA.84.042312}.
\newblock URL \url{https://link.aps.org/doi/10.1103/PhysRevA.84.042312}.

\bibitem[Liu and Zhang(2015)]{small_grover}
Yang Liu and FeiHao Zhang.
\newblock First experimental demonstration of an exact quantum search algorithm
  in nuclear magnetic resonance system.
\newblock \emph{Science China Physics, Mechanics {\&} Astronomy}, 58\penalty0
  (7):\penalty0 1--6, 2015.
\newblock ISSN 1869-1927.
\newblock \doi{10.1007/s11433-015-5661-z}.
\newblock URL \url{http://dx.doi.org/10.1007/s11433-015-5661-z}.

\bibitem[Peruzzo et~al.(2014)Peruzzo, McClean, Shadbolt, Yung, Zhou, Love,
  Aspuru-Guzik, and O’Brien]{quantum_simulation}
Alberto Peruzzo, Jarrod McClean, Peter Shadbolt, Man-Hong Yung, Xiao-Qi Zhou,
  Peter~J. Love, Alán Aspuru-Guzik, and Jeremy~L. O’Brien.
\newblock A variational eigenvalue solver on a photonic quantum processor.
\newblock \emph{Nature Communications}, 5\penalty0 (4213), 2014.
\newblock \doi{10.1038/ncomms5213}.
\newblock URL \url{http://dx.doi.org/10.1038/ncomms5213}.

\bibitem[DiCarlo et~al.(2009)DiCarlo, Chow, Gambetta, Bishop, Johnson,
  Schuster, Majer, Blais, Frunzio, Girvin, and Schoelkopf]{early_qec}
L.~DiCarlo, J.~M. Chow, J.~M. Gambetta, Lev~S. Bishop, B.~R. Johnson, D.~I.
  Schuster, J.~Majer, A.~Blais, L.~Frunzio, S.~M. Girvin, and R.~J. Schoelkopf.
\newblock Demonstration of two-qubit algorithms with a superconducting quantum
  processor.
\newblock \emph{Nature}, 460\penalty0 (7252):\penalty0 240--244, 2009.
\newblock \doi{10.1038/nature08121}.
\newblock URL \url{http://dx.doi.org/10.1038/nature08121}.

\bibitem[{Chamberland} et~al.(2017){Chamberland}, {Iyer}, and
  {Poulin}]{chamberland}
C.~{Chamberland}, P.~{Iyer}, and D.~{Poulin}.
\newblock {Fault-Tolerant Quantum Computing in the Pauli or Clifford Frame with
  Slow Error Diagnostics}.
\newblock \emph{ArXiv e-prints}, April 2017.

\bibitem[Kaveh et~al.(2010)Kaveh, A., and Lorenza]{refocusing}
Khodjasteh Kaveh, Lidar~Daniel A., and Viola Lorenza.
\newblock Arbitrarily accurate dynamical control in open quantum systems.
\newblock \emph{Phys. Rev. Lett.}, 104:\penalty0 090501, Mar 2010.
\newblock \doi{10.1103/PhysRevLett.104.090501}.
\newblock URL \url{https://link.aps.org/doi/10.1103/PhysRevLett.104.090501}.

\bibitem[Motzoi et~al.(2009)Motzoi, Gambetta, Rebentrost, and Wilhelm]{Wilhelm}
F.~Motzoi, J.~M. Gambetta, P.~Rebentrost, and F.~K. Wilhelm.
\newblock Simple pulses for elimination of leakage in weakly nonlinear qubits.
\newblock \emph{Phys. Rev. Lett.}, 103:\penalty0 110501, Sep 2009.
\newblock \doi{10.1103/PhysRevLett.103.110501}.
\newblock URL \url{https://link.aps.org/doi/10.1103/PhysRevLett.103.110501}.

\bibitem[Gottesman(1997)]{gottesman}
Daniel Gottesman.
\newblock \emph{Stabilizer Codes and Quantum Error Correction}.
\newblock Caltech Ph.D. Thesis, 1997.

\bibitem[Lidar and Brun(2013)]{lidar}
Daniel~A. Lidar and Todd~A. Brun.
\newblock \emph{Quantum Error Correction}.
\newblock Cambridge University Press, 2013.

\bibitem[Preskill(1998)]{threshold}
John Preskill.
\newblock Reliable quantum computers.
\newblock \emph{Proceedings of the Royal Society}, A:\penalty0 385--410, 1998.
\newblock \doi{10.1098/rspa.1998.0167}.
\newblock URL \url{https://link.aps.org/doi/10.1098/rspa.1998.0167}.

\bibitem[Kitaev(2003)]{Kitaev2003}
A.Yu. Kitaev.
\newblock Fault-tolerant quantum computation by anyons.
\newblock \emph{Annals of Physics}, 303\penalty0 (1):\penalty0 2 -- 30, 2003.
\newblock ISSN 0003-4916.
\newblock \doi{http://doi.org/10.1016/S0003-4916(02)00018-0}.
\newblock URL
  \url{http://www.sciencedirect.com/science/article/pii/S0003491602000180}.

\bibitem[Freedman and Meyer(2001)]{freedman2001projective}
Michael~H Freedman and David~A Meyer.
\newblock Projective plane and planar quantum codes.
\newblock \emph{Foundations of Computational Mathematics}, 1\penalty0
  (3):\penalty0 325--332, 2001.

\bibitem[Bravyi and Kitaev(1998)]{bravyi1998quantum}
Sergey~B Bravyi and A~Yu Kitaev.
\newblock Quantum codes on a lattice with boundary.
\newblock \emph{arXiv preprint quant-ph/9811052}, 1998.

\bibitem[Bombin and Martin-Delgado(2007)]{bombin2007optimal}
H~Bombin and Miguel~A Martin-Delgado.
\newblock Optimal resources for topological two-dimensional stabilizer codes:
  Comparative study.
\newblock \emph{Physical Review A}, 76\penalty0 (1):\penalty0 012305, 2007.

\bibitem[Dennis et~al.(2002)Dennis, Kitaev, Landahl, and Preskill]{dklp}
Eric Dennis, Alexei Kitaev, Andrew Landahl, and John Preskill.
\newblock Topological quantum memory.
\newblock \emph{Journal of Mathematical Physics}, 43\penalty0 (9):\penalty0
  4452--4505, 2002.
\newblock \doi{10.1063/1.1499754}.
\newblock URL \url{http://dx.doi.org/10.1063/1.1499754}.

\bibitem[Fowler et~al.(2009)Fowler, Stephens, and Groszkowski]{fowler2009high}
Austin~G Fowler, Ashley~M Stephens, and Peter Groszkowski.
\newblock High-threshold universal quantum computation on the surface code.
\newblock \emph{Physical Review A}, 80\penalty0 (5):\penalty0 052312, 2009.

\bibitem[Tomita and Svore(2014)]{tomita}
Yu~Tomita and Krysta~M. Svore.
\newblock Low-distance surface codes under realistic quantum noise.
\newblock \emph{Phys. Rev. A}, 90:\penalty0 062320, Dec 2014.
\newblock \doi{10.1103/PhysRevA.90.062320}.
\newblock URL \url{https://link.aps.org/doi/10.1103/PhysRevA.90.062320}.

\bibitem[Schrijver(2003)]{schrijver2003combinatorial}
A.~Schrijver.
\newblock \emph{Combinatorial Optimization: Polyhedra and Efficiency}.
\newblock Number v. 1 in Algorithms and Combinatorics. Springer, 2003.
\newblock ISBN 9783540443896.
\newblock URL \url{https://books.google.nl/books?id=mqGeSQ6dJycC}.

\bibitem[Edmonds(1965)]{edmonds}
Jack Edmonds.
\newblock Paths, trees, and flowers.
\newblock \emph{Canadian Journal of Mathematics}, 17:\penalty0 449--467, 1965.
\newblock \doi{10.4153/CJM-1965-045-4}.
\newblock URL \url{http://dx.doi.org/10.4153/CJM-1965-045-4}.

\bibitem[Kolmogorov(2009)]{kolmogorov}
Vladimir Kolmogorov.
\newblock Blossom v: a new implementation of a minimum cost perfect matching
  algorithm.
\newblock \emph{Mathematical Programming Computation}, 1:\penalty0 43--67,
  2009.
\newblock \doi{10.1007/s12532-009-0002-8}.
\newblock URL \url{http://dx.doi.org/10.1007/s12532-009-0002-8}.

\bibitem[O'Brien et~al.(2017)O'Brien, Tarasinski, and
  DiCarlo]{toms_fancy_simulation_paper}
TE~O'Brien, B~Tarasinski, and L~DiCarlo.
\newblock Density-matrix simulation of small surface codes under current and
  projected experimental noise.
\newblock \emph{arXiv preprint arXiv:1703.04136}, 2017.

\bibitem[Versluis et~al.(2016)Versluis, Poletto, Khammassi, Haider, Michalak,
  Bruno, Bertels, and DiCarlo]{leos_fancy_cycle_paper}
R~Versluis, S~Poletto, N~Khammassi, N~Haider, DJ~Michalak, A~Bruno, K~Bertels,
  and L~DiCarlo.
\newblock Scalable quantum circuit and control for a superconducting surface
  code.
\newblock \emph{arXiv preprint arXiv:1612.08208}, 2016.

\bibitem[Fowler(2015)]{fowler_o1}
Austin~G. Fowler.
\newblock Minimum weight perfect matching of fault-tolerant topological quantum
  error correction in average $o(1)$ parallel time.
\newblock \emph{Quantum Information \& Computation}, 15:\penalty0 145--158,
  2015.

\bibitem[Torlai and Melko(2016)]{torlai}
Giacomo Torlai and Roger~G Melko.
\newblock A neural decoder for topological codes.
\newblock \emph{arXiv preprint arXiv:1610.04238}, 2016.

\bibitem[Poulin(2006)]{pure_errors}
David Poulin.
\newblock Optimal and efficient decoding of concatenated quantum block codes.
\newblock \emph{Phys. Rev. A}, 74:\penalty0 052333, Nov 2006.
\newblock \doi{10.1103/PhysRevA.74.052333}.
\newblock URL \url{https://link.aps.org/doi/10.1103/PhysRevA.74.052333}.

\bibitem[MacKay(2003)]{ml_textbook}
David~JC MacKay.
\newblock \emph{Information theory, inference and learning algorithms}.
\newblock Cambridge university press, 2003.

\bibitem[Abadi et~al.(2015)Abadi, Agarwal, Barham, Brevdo, Chen, Citro,
  Corrado, Davis, Dean, Devin, Ghemawat, Goodfellow, Harp, Irving, Isard, Jia,
  Jozefowicz, Kaiser, Kudlur, Levenberg, Man\'{e}, Monga, Moore, Murray, Olah,
  Schuster, Shlens, Steiner, Sutskever, Talwar, Tucker, Vanhoucke, Vasudevan,
  Vi\'{e}gas, Vinyals, Warden, Wattenberg, Wicke, Yu, and Zheng]{tensorflow}
Mart\'{\i}n Abadi, Ashish Agarwal, Paul Barham, Eugene Brevdo, Zhifeng Chen,
  Craig Citro, Greg~S. Corrado, Andy Davis, Jeffrey Dean, Matthieu Devin,
  Sanjay Ghemawat, Ian Goodfellow, Andrew Harp, Geoffrey Irving, Michael Isard,
  Yangqing Jia, Rafal Jozefowicz, Lukasz Kaiser, Manjunath Kudlur, Josh
  Levenberg, Dan Man\'{e}, Rajat Monga, Sherry Moore, Derek Murray, Chris Olah,
  Mike Schuster, Jonathon Shlens, Benoit Steiner, Ilya Sutskever, Kunal Talwar,
  Paul Tucker, Vincent Vanhoucke, Vijay Vasudevan, Fernanda Vi\'{e}gas, Oriol
  Vinyals, Pete Warden, Martin Wattenberg, Martin Wicke, Yuan Yu, and Xiaoqiang
  Zheng.
\newblock {TensorFlow}: Large-scale machine learning on heterogeneous systems,
  2015.
\newblock URL \url{http://tensorflow.org/}.
\newblock Software available from tensorflow.org.

\bibitem[Mehta(2012)]{clock_cycle}
Nick Mehta.
\newblock Xilinx redefines power, performance, and design productivity with
  three innovative 28 nm fpga families: Virtex-7, kintex-7, and artix-7
  devices, 2012.
\newblock URL
  \url{xilinx.com/support/documentation/white_papers/wp373_V7_K7_A7_Devices.pdf}.
\newblock www.xilinx.com.

\bibitem[Bombin and Martin-Delgado(2006)]{color_code_defn}
Hector Bombin and Miguel~Angel Martin-Delgado.
\newblock Topological quantum distillation.
\newblock \emph{Physical review letters}, 97\penalty0 (18):\penalty0 180501,
  2006.

\bibitem[Landahl et~al.(2011)Landahl, Anderson, and Rice]{color_code_ft}
Andrew~J Landahl, Jonas~T Anderson, and Patrick~R Rice.
\newblock Fault-tolerant quantum computing with color codes.
\newblock \emph{arXiv preprint arXiv:1108.5738}, 2011.

\bibitem[Bombin et~al.(2012)Bombin, Duclos-Cianci, and Poulin]{color_code_rg}
Hector Bombin, Guillaume Duclos-Cianci, and David Poulin.
\newblock Universal topological phase of two-dimensional stabilizer codes.
\newblock \emph{New Journal of Physics}, 14\penalty0 (7):\penalty0 073048,
  2012.

\end{thebibliography}
\end{document}